\documentclass[twocolumn,floats,prl,aps,showpacs]{revtex4}
\usepackage{graphicx}
\usepackage{amsfonts}
\usepackage{amsmath}
\usepackage{amssymb}
\usepackage{exscale}
\usepackage{color}

\begin{document}

\title{Effects of fermion exchanges on the polarization of exciton condensates}
\author{ Monique Combescot$^1$, Roland Combescot$^{2,3}$, Mathieu Alloing$^{4,1}$,  Fran\c{c}ois Dubin$^{1,4}$}
\affiliation{(1)Institut des NanoSciences de Paris, Universit\'e Pierre et Marie Curie,
CNRS, Tour 22, 4 place Jussieu, 75005 Paris}
\affiliation{(2) Laboratoire de Physique Statistique, Ecole Normale Sup\'{e}rieure, UPMC  
Paris 06, Universit\'e Paris Diderot, CNRS, 24 rue Lhomond, 75005 Paris,
France}
\affiliation{(3) Institut Universitaire de France, 103 boulevard Saint-Michel, 75005 Paris, France}
\affiliation{(4) ICFO-The Institute of Photonic Sciences, 3 Av. Carl Friedrich Gauss, 08860
Castelldefels (Barcelona), Spain}

%\pacs{73.63.Hs}{Quantum wells}
%\pacs{78.47.jd}{Time resolved luminescence}
%\pacs{03.75.Hh}{Static properties of condensates}
\date{\today }

\begin{abstract}
Exchange processes are responsible for the stability of \textit{elementary} boson condensates with respect to their possible fragmentation. This remains true for \textit{composite} bosons when single fermion exchanges are included but spin degrees of freedom are ignored. We here show that their inclusion can produce a "spin-fragmentation" of a condensate of dark excitons, \textit{i.e.}, an unpolarized condensate with equal amount of dark excitons with spins $(+2)$ and $(-2)$. Quite surprisingly, for spatially indirect excitons of semiconductor bilayers, we predict that the condensate polarization can switch from unpolarized to fully polarized, depending on the distance between the layers confining electrons and holes. Remarkably, the threshold distance associated to this switching lies in the regime where experiments are nowadays carried out.  
\end{abstract}
%\pacs{03.75.Hh}

\maketitle
 
Free elementary bosons are well known to undergo a quantum phase transition with a macroscopic amount of bosons condensed in the same ground
state when their number gets larger than $N_c(T)\propto(L/\lambda_T)^D $ where $D$ is the space dimension, $L$ the sample size and $\lambda_T$ the
thermal de Broglie wave length defined by $k_B T=\hbar^2/2m\lambda_T^2$. Yet it is surprising that, in a macroscopic sample, the $\textbf{k}=\textbf{0}$ ground state plays such an important role compared to states with very small momentum because, in a size $L$ sample, the energy difference between $N$ free bosons with momentum $k=2\pi/L$ and with momentum $k=0$ scales as $N/L^2$ which is underextensive. 

Nozi\`eres \cite{noz} has pointed out that, when interactions between bosonic particles are included, 
exchange processes between the two components of the fragmented state ($\textbf{k},-\textbf{k}$) produce an extensive energy difference. So, Bose-Einstein condensation with all bosons in the same $\textbf{k}=\textbf{0}$ state is not driven by kinetic energy difference but by exchange interaction processes. For composite bosons made of two fermions \cite{PhysReport}, for example semiconductor excitons made of an electron-hole pair, we have shown \cite{comb-snoke} that this conclusion remains true. This result is not obvious at first because, in addition to exchanging simultaneously the two fermions of a composite boson - which amounts to take the composite boson as a whole - one also has to consider exchanges of only one of the two fermions. 

In this letter we consider the role played by fermion exchanges on the spin degrees of freedom of exciton condensates. We show that they lead to a possible variation of the spin structure of a condensate from "spin-unfragmented",\emph{ i.e.}, a fully polarized state with all excitons having same spin, to "spin-fragmented",\emph{ i.e.}, unpolarized with an equal amount of up-spin and down-spin excitons. We actually show that the latter realisation is quasi-degenerate with a linearly polarized condensate in the thermodynamic limit. Finally,
we consider indirect excitons \cite{lozovik} which have recently lead to remarkable observations \cite{Butov_2012,Timofeev_2012,Butov_2012_b,Rapaport_2013,Holleitner_2013,Stern_2013,Butov_2013,Santos_2014,francois}. For these, we predict that depending on the distance between the quantum wells confining electrons and holes, i.e. depending on the excitons electric dipole moment,  the spin structure of a condensate may switch from unpolarized to fully polarized. Quite interestingly, the threshold distance associated to this switch of the condensate polarization lies in the regime where current experimental studies are carried out. 

\textbf{Simplified approach}:  In order to find the exciton condensate polarization induced by interactions in an easy way, let us start with a simplified approach.

\textbf{(i)} 
It is useful to briefly recall Nozi\`eres's argument on the "fragmentation" of  Bose-Einstein condensates \cite{noz}. We consider the simple Hamiltonian with contact interactions
\begin{eqnarray}\label{}
{\bar H}{=} \sum_{\textbf{q}} \epsilon_\textbf{q} {\bar B}^{\dag}_\textbf{q} {\bar B}_\textbf{q}{+} \frac{V}{2}  \sum_{\textbf{q}_i}{\bar B}^{\dag}_{\textbf{q}_4}{\bar B}^{\dag}_{\textbf{q}_3}
{\bar B}_{\textbf{q}_2}{\bar B}_{\textbf{q}_1} \delta_{\textbf{q}_4{+}\textbf{q}_3,\textbf{q}_1{+}\textbf{q}_2}
\end{eqnarray}
and look for the corresponding energy in the fragmentated state $|{\bar \Phi}_{N,N'}\rangle{=}(\bar{B}^\dag_\textbf{k})^N(\bar{B}^\dag_{-\textbf{k}})^{N'}|0\rangle$. Iteration of 
$\left[{\bar B}_{\textbf{q}},{\bar B}^{\dag}_{\textbf{q}'}\right]{=}\delta_{\textbf{q},\textbf{q}'}$ gives 
 $\langle {\bar \Phi}_{N,N'}|{\bar \Phi}_{N,N'}\rangle{=}N!  N'! $
The $\bar H$ mean value in the normalized $|{\bar \Phi}_{N,N'}\rangle$ state is found as
\begin{eqnarray}\label{eq5}
\langle {\bar H{-}\epsilon_\textbf{k}(N{+}N')}\rangle_{N,N'}{=}
\frac{V}{2}\Big\{N(N{-}1){+}N'(N'{-}1){+}4NN'\Big\}
\end{eqnarray}
the $N N'$ term coming from exchange processes between ($\textbf{k}$) and ($\textbf{-k}$) bosons. So, the energy difference between fragmented and fully condensed states reads as
\begin{eqnarray}\label{}
\langle {\bar H}\rangle_{N,N}-\langle {\bar H}\rangle_{2N,0}=V N^2
\end{eqnarray} 
For $V$ positive, as necessary to avoid collapse,
this difference is positive and extensive since $V$ scales as one over the sample volume. A similar result is found \cite{tony} for the coherent state $|\bar \Phi_{2N}^{(L)}\rangle=(B^{\dag}_\textbf{k}+B^{\dag}_\textbf{-k})^{2N}|0\rangle$.
Equation (2) also shows that 
the kinetic energy difference between the fragmented state $|{\bar \Phi}_{N,N}\rangle$ taken for $k=2\pi /L$ and the condensed state $|{\bar \Phi}_{2N,0}\rangle$ taken for $k=0$ scales as $N/L^2$. Hence it is underextensive and negligible in the thermodynamic limit. This shows that Bose-Einstein condensation with all bosons in the {\it same} $\textbf{k}=\textbf{0}$ state is driven by interactions, not by kinetic energy.

\textbf{(ii)}
To study the condensate polarization, we consider two kinds of elementary bosons having up or down spin and same energy (taken as zero), their creation operators being $\bar{B}^{\dag}_{+}$ or $\bar{B}^{\dag}_{-}$ respectively. It is convenient to introduce the scattering $W$ between
same-spin bosons and the scattering $V$ which is independent of the boson spins, and to write the corresponding Hamiltonian as
\begin{eqnarray}\label{}
{\bar H_{pol}} =  \frac{V}{2} \sum_{s=\pm}  \sum_{s'=\pm}{\bar B}^{\dag}_{s}{\bar B}^{\dag}_{s'}{\bar B}_{s'}{\bar B}_{s}
+\frac{W}{2}  \sum_{s=\pm}{\bar B}^{\dag}_{s}{\bar B}^{\dag}_{s}{\bar B}_{s}{\bar B}_{s}
\end{eqnarray}
A calculation similar to the previous one gives the energy in the state $|\bar\Psi_{N,N'}\rangle=
(\bar{B}^{\dag}_{+})^N(\bar{B}^{\dag}_{-})^{N'}|0\rangle$ as
\begin{eqnarray}\label{}
\langle {\bar H_{pol}}\rangle_{N,N'}{=}
\frac{V}{2}\Big\{N(N-1)+N'(N'-1)+2NN' \Big\}
\\ \nonumber
{+}\frac{W}{2}{\Big\{}N(N{-}1){+}N'(N'{-}1)\Big\}
\end{eqnarray}
For a given ($N+N'$), the first bracket, equal to $(N+N')(N+N'-1)$, does not depend on polarization as reasonable since $V$ acts between arbitrary spins. By contrast, same-spin scatterings lead to an energy difference between the unpolarized state $|\bar\Psi_{N,N}\rangle$ and the fully polarized state $|\bar\Psi_{2N,0}\rangle$ which reads as
\begin{eqnarray}\label{}
\langle {\bar H_{pol}}\rangle_{N,N}-\langle {\bar H_{pol}}\rangle_{2N,0}
=-W N^2
\end{eqnarray}
So the lowest energy state is the unpolarized state $|\bar\Psi_{N,N}\rangle$ for $W>0$, while for $W<0$, it is the polarized state
${|\bar\Psi}_{2N,0}\rangle$, degenerate with ${|\bar\Psi}_{0,2N}\rangle$.

We can also consider the linearly polarized state 
$|\bar\Psi^{(L)}_{2N}\rangle=(\bar{C}^{\dag}_{+})^{2N}|0\rangle$ with
$\bar{C}^{\dag}_{\pm}=(\bar{B}^{\dag}_{+} \pm \bar{B}^{\dag}_{-})/\sqrt{2}$.
By writing ${\bar H_{pol}} $
in terms of $\bar{C}^{\dag}_{\pm}$, it is easy to show that $\langle {\bar H_{pol}}\rangle^{(L)}_{2N}=(V/2+W/4)2N(2N-1)$ from which we get $\langle {\bar H_{pol}}\rangle^{(L)}_{2N}-\langle {\bar H_{pol}}\rangle_{N,N}=WN/2$. So, for $W>0$, the lowest energy state indeed is the unpolarized state but, as $WN$ is underextensive, the energy difference with the linearly polarized state goes to zero in the thermodynamic limit - the polarized state ${|\bar\Psi}_{2N,0}\rangle$ still having the lowest energy for $W<0$.

\textbf{Composite boson approach}:  In the simplified approach, the exciton composite nature appears through the existence of spin dependent scatterings, this dependence coming from the fact that, in Coulomb processes, electrons and holes conserve their own spin. Otherwise, excitons are taken as elementary bosons because their "deviation operators"
$D_{mi}=[B_m,B^{\dag}_i]-\delta_{m,i}$ 
are not considered. It is commonly said that the $D_{mi}$ operators can be neglected at small density - which is not correct. Indeed, the "moth-eaten effect"
induced by the $D_{mi}$'s shows up, in particular, in
\begin{eqnarray}\label{}
\langle 0 | {B}_{0}^N B^{\dag N}_0|0\rangle = N ! F_N
\end{eqnarray}
where $F_N$ is not of the order of 1, but is exponentially small \cite{Tanguy}: it decreases with exciton number as $e^{-N\eta}$ where 
the gas parameter $\eta=N(a_X/L)^D$ measures the exciton density at the scale of the Bohr radius $a_X$. However this huge "moth-eaten effect" does not have dramatic consequences because $F_N$ ultimately appears in physical quantities through ratios like 
$F_{N-1}/F_N\simeq1+\eta  f_D$
with $f_D = 4\pi /5$ in 2D. Accordingly, when looking for density effects, these $F_N$ factors must be kept. This implies
to keep the $D_{mi}$ operators from the very beginning which is what the many-body formalism \cite{PhysReport} for composite
bosons enables us to do. 

From dimensional arguments, Coulomb scatterings involving two excitons scale as $R_X(a_X/L)^D$ while those involving
($n+1$) excitons scale as $R_X(a_X/L)^{nD}$; they appear with a $N^{n+1}$ prefactor which comes from the 
$\left(_{n+1}^{N}\right)$ ways to select $(n+1)$ excitons among $N$, bringing contribution in $N\,R_X\,\eta^n$. So, effects at first order in density, 
to which we restrict ourselves in this paper, come from processes involving two excitons among $N$.

Dark excitons with spin $(\pm2)$ have a lower energy than bright excitons $(\pm1)$ for the very same reason that they are not coupled to light \cite{Michael}. 
 So, excitons condense in a dark state \cite{CBC}. This  leads us to study the polarization of the condensate through the mean value of the semiconductor Hamiltonian $H$ in the partially polarized dark exciton state
$|\Phi_{N,N'}\rangle=(B^{\dag}_{0,2})^N(B^{\dag}_{0,-2})^{N'}|0\rangle$
\begin{eqnarray}\label{}
\Delta_{N,N'}= \frac{\langle \Phi_{N,N'}|H-(N+N')E_0|\Phi_{N,N'}\rangle}{\langle \Phi_{N,N'}|\Phi_{N,N'}\rangle}
\end{eqnarray}
where we have subtracted the energy $E_0$ corresponding to the isolated dark exciton in its ground state. In the coherent linearly polarized state
$|\Psi_{N}^{(L)}\rangle=(B^{\dag}_{0,2}+B^{\dag}_{0,-2})^{N}|0\rangle$, $\Delta^{(L)}_{N}$ reads as $\Delta_{N,N'}$ with $|\Phi_{N,N'}\rangle$ replaced by $|\Psi_{N}^{(L)}\rangle$.

The creation operator of a 2D dark exciton $i$ with center-of-mass momentum $\textbf{Q}_i$, relative motion index $\nu_i$ and total
spin ($\pm 2$) reads in terms of electron and hole creation operators 
$a^{\dagger}_{\textbf{p}}$ and 
$b^{\dagger}_{\textbf{p}}$ as
\begin{eqnarray}\label{}
    B^{\dagger}_{i;\pm2}=\sum_{\textbf{p}}a^{\dagger}_{\textbf{p}+\gamma_{e}\textbf{Q}_{i};\pm1/2}b^{\dagger}_{-\textbf{p}+\gamma_{h}\textbf{Q}_{i};\pm3/2} \langle\textbf{p} | \nu_{i}\rangle
\end{eqnarray}
where $\gamma_{e}=1-\gamma_{h}=m_{e}/(m_{e}+m_{h})$. The index $0$ in $B^{\dag}_{0,\pm 2}$
stands for the dark exciton ground state ($\textbf{Q}=\textbf{0}, \nu=\nu_0$) with energy $E_0$.

\emph{(i) Two excitons}

To analyze carrier exchanges in a simple way, let us start with two excitons and compare the energy
of the fully polarized state $|\Phi_{2,0}\rangle$, the unpolarized state $|\Phi_{1,1}\rangle$ and the linearly polarized state
$|\Psi_{2}^{(L)}\rangle$, a rule of the thumb giving the result for $N$ by putting $N(N-1)/2$ in front of the interaction term. The key commutators of the composite boson many-body formalism \cite{PhysReport} give
\begin{eqnarray}\label{}
\hspace{-3mm}\Delta_{1,1}{=} \frac{\sum_{mn}\langle 0|B_{0,-2}B_{0,2}B^{\dagger}_{m,2}B^{\dagger}_{n,-2}|0\rangle\xi(_{m0}^{n 0})}
{\langle 0|B_{0,-2}B_{0,2}B^{\dagger}_{0,2}B^{\dagger}_{0,-2}|0\rangle}= \xi(_{0 0}^{0 0})
\end{eqnarray}
\begin{eqnarray}\label{eq11}
\Delta_{2,0}= \frac{2\xi(_{0 0}^{0 0})-2\xi^{exch} (_{0 0}^{0 0})}{2-2\lambda (_{0 0}^{0 0})}
\end{eqnarray}
From $|\Psi_{2}^{(L)}\rangle=|\Phi_{2,0}\rangle+|\Phi_{0,2}\rangle+2|\Phi_{1,1}\rangle$, this yields
\begin{eqnarray}\label{eq12}
\Delta^{(L)}_{2}= \frac{8\xi(_{0 0}^{0 0})-4\xi^{exch} (_{0 0}^{0 0})}{8-4\lambda (_{0 0}^{0 0})}
\end{eqnarray}
The denominators in Eq.(11-12) come from the norm of the state, calculated through
$[D_{mi},B^{\dagger}_{j}]= \sum_{n}(\lambda (_{m i}^{n j})+\lambda (_{n i}^{m j}))B^{\dagger}_{n}$,
which can be seen as a definition for $\lambda (_{m i}^{n j})$. 
Exchange and direct Coulomb scatterings are related by $\xi^{exch}(_{m i}^{n j})=\sum_{p,q}\lambda (_{m p}^{n q})\xi(_{p i}^{q j})$ with $[V^{\dagger}_{i},B^{\dagger}_{j}]= \sum_{m,n}\xi(_{m i}^{n j})B^{\dagger}_{m}B^{\dagger}_{n}$ and $V^{\dagger}_{i}=[H,B^{\dagger}_{i}]- E_iB^{\dagger}_{i}$.

As shown below, $\xi\geqslant0$ while $\xi^{exch}$ can be
positive or negative. Since $(\xi,\xi^{exch},\lambda)$ scale as $(a_X/L)^D$, which is small in large sample, the above results show that
the lowest energy state is the unpolarized state $|\Phi_{1,1}\rangle$
for $-\xi^{exch}>0$ while it is the fully polarized state $|\Phi_{2,0}\rangle$ or $|\Phi_{0,2}\rangle$
for $-\xi^{exch}<0$. As we show now, this result remains true for $N\gg2$, except that the energy difference between unpolarized and linearly polarized states turns underextensive. 

\emph{(ii) Large exciton number}

Exchange processes between
$N$ excitons having same spin changes their normalization factor from $N !$ to $N ! F_N$. By contrast, exchange
processes between opposite spin dark excitons bring them to the higher energy bright subspace (see Fig.1); so, they should not be considered here.
 Since $B^{\dag}_{0,2}$ and $B^{\dag}_{0,-2}$ are made of different carriers, ${B}_{0,2} B^{\dag}_{0,-2}|0\rangle =0$ which gives
\begin{eqnarray}\label{}
\langle \Phi_{N,N'} | \Phi_{N,N'} \rangle = (N ! F_N)(  N' ! F_{N'})
\end{eqnarray}
By only keeping Coulomb processes involving two excitons among $N+N'$, we get
\begin{eqnarray}\label{}
\langle \Phi_{N,N'} | H-(N+N')E_0 |\Phi_{N,N'} \rangle \hspace{25mm}
\\ \nonumber
\simeq \xi \Big(X_{N,N'}{+}X_{N',N}{+}Y_{N,N'}\Big){-}\xi^{exch}
\Big(X_{N,N'}{+}X_{N',N}\Big)
\end{eqnarray}
where from now on the 0 indices in $\xi$ and $\xi^{exch}$ are omitted for brievity.
$X_{N,N'}$ and $Y_{N',N}$ are defined below.
Direct \textit{and} exchange Coulomb processes exist between the $N$ excitons $(+2)$, or between the $N'$ excitons $(-2)$. They bring a factor $X_{N,N'}+X_{N',N}$ with
\begin{eqnarray}\label{}
X_{N,N'}{=}\frac{N^2(N{-}1)^2}{2}\Big\{(N{-}2)!F_{N{-}2}\Big\}\Big\{(N')!F_{N'} \Big\}
\end{eqnarray}
the curly brackets coming from the unaffected $(N-2)$ excitons $(+2)$ and $N'$ excitons $(-2)$. The $N$ excitons $(+2)$ also have direct Coulomb processes, but no exchange, with the $N'$ excitons $(-2)$. They bring a factor 
\begin{eqnarray}\label{}
Y_{N,N'}{=}N^2N'^2\Big\{(N{-}1)!F_{N{-}1}\Big\}\Big\{N'-1)!F_{N'-1} \Big\}
\end{eqnarray}

This gives for a fully polarized condensate
\begin{eqnarray}\label{eq25}
\Delta_{2N,0}=
\frac{2N(2N-1)}{2}
\frac{F_{2N-2}}{F_{2N}}
 \Big(\xi-\xi^{exch}\Big) 
\end{eqnarray}
 the $F_N$ ratio reducing to $1$ at lowest order in density. For an unpolarized condensate, we get
\begin{eqnarray}\label{eq26}
\Delta_{N,N}{=}\frac{2N(N{-}1)}{2}
\frac{F_{N-2}}{F_{N}}
 \Big(\xi{-}\xi^{exch} \Big) 
 {+}N^2\Big(\frac{F_{N-1}}{F_{N}}\Big)^2\xi\\ \nonumber
\simeq N(2N-1) \xi- N(N-1)\xi^{exch}
\end{eqnarray}

$\Delta_{N,N}$ and $\Delta_{2N,0}$ have the same amount of direct Coulomb processes, as reasonable because  these processes do not depend on spin, while $\Delta_{N,N}$ has less exchange Coulomb processes, $\Delta_{2N,0}-\Delta_{N,N}\simeq-N^2\xi^{exch}$. So, the
unpolarized state $ | \Phi_{N,N} \rangle$ has the lowest energy for $-\xi^{exch}>0$, but the fully
polarized state $ | \Phi_{2N,0} \rangle$,  degenerate with $ | \Phi_{0,2N} \rangle$, becomes the lowest state for $-\xi^{exch}<0$. 

Let us also consider the linearly polarized state
\begin{eqnarray}\label{}
|\Psi_{2N}^{(L)}\rangle=(B^{\dag}_{0,2}+B^{\dag}_{0,-2})^{2N}|0\rangle=  \sum_{p=0}^{2N} \left(_{\,p}^{2N}\right)| \Phi_{2N-p,p} \rangle
\end{eqnarray}
The calculation of the Hamiltonian mean value is more demanding due to the "moth-eaten effect" between different numbers of (+2) and (-2) excitons. Indeed,
Eq.(13) becomes $\langle \Psi_{2N}^{(L)} |\Psi_{2N}^{(L)}\rangle=(2N)! F_{2N}^{(L)}$ with 
$F_{2N}^{(L)}$ given by
\begin{eqnarray}\label{}
F_{2N}^{(L)}= \sum_{p=0}^{2N}\left(_{\,p}^{2N}\right) F_{2N-p}F_p
\end{eqnarray}
and Eq.(14) becomes
\begin{eqnarray}\label{}
\langle \Psi_{2N}^{(L)} | H-2N E_0 |\Psi_{2N}^{(L)} \rangle\!=\!(2N)!\bigg\{\!\xi{\mathcal F}_{2N}-\xi^{exch}{\mathcal G}_{2N}\!\bigg\}
\end{eqnarray}
with ${\mathcal F}_{2N}$ given by
\begin{eqnarray}\label{F2N}
{\mathcal F}_{2N}=
\sum_{p=0}^{2N} \left(_{\,p}^{2N}\right) \bigg\{\left(_{\,2}^{2N-p}\right) F_p F_{2N-p-2}  \hspace{20mm}\\ \nonumber
+\left(_{\,1}^{2N-p}\right) \left(_{\,1}^{p}\right)\,F_{p-1} F_{2N-p-1}+\left(_{\,2}^{p}\right) F_{p-2} F_{2N-p} \bigg\}
\end{eqnarray}
${\mathcal G}_{2N}$ reads as ${\mathcal F}_{2N}$ without the $F_{p-1} F_{2N-p-1}$ term which, like $Y_{N,N'}$, comes from processes involving 
\textit{one} exciton $(+2)$ and \textit{one} exciton $(-2)$ for which no exchange can occur.

For $N=2$, these results agree with Eqs.(11,12). To get them for large $N$, we use the fact that the binomial coefficients $\left(_{\,p}^{2N}\right)$ 
are very much peaked on $p=N$ while
\begin{eqnarray}\label{}
\left(_{\,2}^{2N-p}\right)+\left(_{\,1}^{2N-p}\right) \left(_{\,1}^{p}\right)+\left(_{\,2}^{p}\right)=\left(_{\,2}^{2N}\right)
\end{eqnarray}
As $F_N$ scales as $e^{-N\eta}$ while
$F_{N-1}/F_N=1+{\mathcal O}(\eta)$, 
we get
\begin{eqnarray}\label{}
\Delta^{(L)}_{2N} \simeq
\left(_{\,2}^{2N}\right)\xi -2\left(_{\,2}^{N}\right)\xi^{exch}\simeq\Delta_{N,N}
\end{eqnarray}
within underextensive terms.
So, for $-\xi^{exch} >0$, the unpolarized state $|\Phi_{N,N}\rangle$, quasi-degenerate with the linearly polarized state 
$|\Psi_{2N}^{(L)}\rangle$, has the lowest energy while, for $-\xi^{exch}<0$, the exciton condensate is fully polarized.

Let us note that, in this study, we have eliminated Coulomb exchange processes between opposite spin dark excitons because they transfer them into the higher energy bright subspace. As the bright/dark splitting $\delta_{BD}$ is independent of sample volume, the energy increase $N\delta_{BD}$ for $N$ excitons turning bright is extensive by contrast with $N$ excitons having a $2\pi/L$ momentum. So, in macroscopic samples, no additional interactions are required to drive exciton condensation toward lowest energy dark states. Yet, when exciton density increases, exchange Coulomb processes cannot be neglected anymore.
We have shown \cite{Roland_2012} that, above a density threshold, they generate a bright component to the exciton condensate which makes it "gray" and possible to study through the luminescence of its bright part.

\textbf{Dipolar excitons}: To study Bose-Einstein condensation of excitons, spatially indirect (or "dipolar") excitons are among the most promising candidates. These are engineered by imposing a spatial separation between electrons and holes, e.g. in a double quantum well \cite{lozovik}. Recently, it was shown at sub-Kelvin temperatures that dipolar excitons can exhibit a quantum statistical distribution marked by a dominant fraction of dark excitons \cite{Rapaport_2013,francois}. Macroscopic spatial coherence \cite{Butov_2012,Butov_2012_b,francois} and spontaneous spin polarisation \cite{Butov_2012,francois} of bright excitons have also been observed. 

\begin{figure}
\centering
%\hspace*{20mm}
\includegraphics[width=80mm]{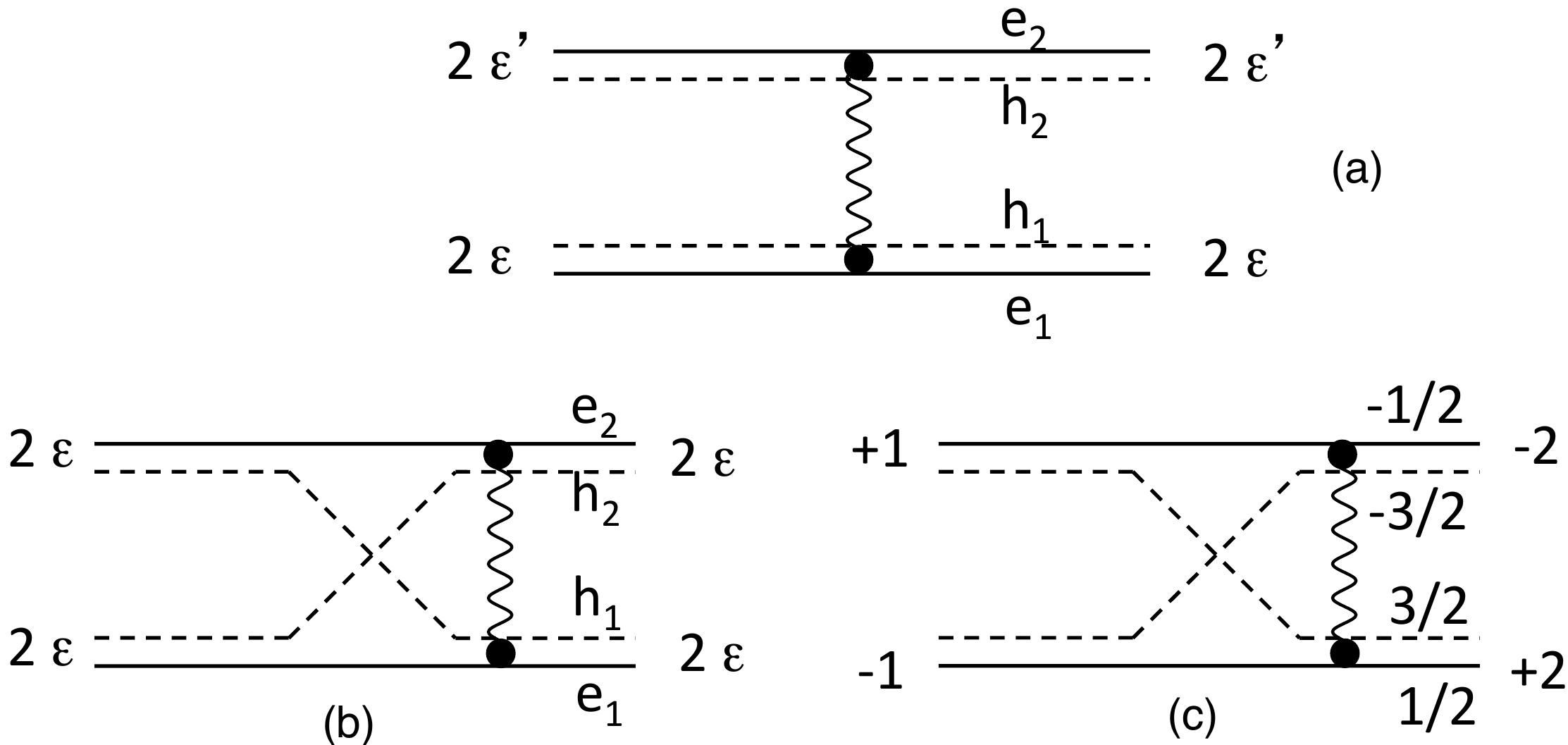}
\caption{a) Shiva diagrams for direct Coulomb scatterings between dark excitons. Exchange Coulomb scatterings between same spin (b) and opposite spin (c) dark excitons. $(\epsilon ,\epsilon ')= \pm 1$.}
\label{Fig1}
\end{figure}

To mimic these dipolar excitons, scattering between arbitrary-spin excitons must be associated with direct Coulomb processes,
$V=\xi$, and scattering between same-spin excitons to exchange Coulomb processes, 
\textit{i.e.}, 
$W=-\xi^{exch}$. For carrier coordinates $({\bf r}_e,d)$
and
$({\bf r}_h,0)$, with $d$ being the distance between electron and hole planes and ${\bf r}_{e,h}$ being 2D vectors along the plane, direct Coulomb scattering between two dark excitons in their ground state $(\nu=\nu_0,{\bf Q}={\bf 0})$, read from Fig.1a, is given by \cite{PhysReport}
\begin{eqnarray}\label{eq1}
\xi_d{=}\!\!\int \{d {\bf r} \} \Big|\varphi_{\nu_0}(\textbf{d}{+}{\bf r}_{e_1}{-}{\bf r}_{h_1})
\Big|^2\Big|\varphi_{\nu_0}(\textbf{d}{+}{\bf r}_{e_2}{-}{\bf r}_{h_2})\Big|^2 
\hspace{7mm}
\\ \nonumber
\times\left[\frac{e^2}{|{\bf r}_{e_1}{-}{\bf r}_{e_2|}}{+}\frac{e^2}{|{\bf r}_{h_1}{-}{\bf r}_{h_2|}}{-}\frac{e^2}{|\textbf{d}{+}{\bf r}_{e_1}{-}{\bf r}_{h_2}|}
{-}\frac{e^2}{|\textbf{d}{+}{\bf r}_{e_2}{-}{\bf r}_{h_1}|}\right]
\end{eqnarray}
where $\varphi_{\nu_0}$ is the relative motion wave function of the exciton ground state. $\xi_d$, which cancels for $d=0$ as seen by exchanging $({\bf r}_{e_1}$, ${\bf r}_{h_1})$, stays positive for finite $d$.

The exchange Coulomb scattering $\xi^{exch}_d$ between two same-spin dark excitons, shown in Fig.1b, reads as $\xi_d$ except for the wave function part which is replaced by
\begin{eqnarray}\label{}
\varphi^*_{\nu_0}({\textbf{d}+\bf r}_{e_1}-{\bf r}_{h_2})\varphi^*_{\nu_0}(\textbf{d}+{\bf r}_{e_2}-{\bf r}_{h_1})\hspace{15mm} \\ \nonumber
\times \varphi_{\nu_0}(\textbf{d}+{\bf r}_{e_1}-{\bf r}_{h_1})\varphi_{\nu_0}(\textbf{d}+{\bf r}_{e_2}-{\bf r}_{h_2})
\end{eqnarray}
We have shown \cite{PhysReport,dupertuis} that $\xi^{exch}_{d=0}=\xi_D\,R_X\,(a_X/L)^D$
with $(R_X,a_X)$ being the $3D$ exciton Rydberg and Bohr radius and $\xi_D$ a numerical factor equal to
$-(8\pi -315\pi ^3/512) \simeq -6.057$ in $2D$. When $d$ increases, the electron-hole part of the bracket in Eq.(\ref{eq1}) goes to zero. Since
the wave function part stays positive, $\xi^{exch}_d$ becomes positive for $d$ larger than a threshold value $d^*$.
We find numerically $d^*\simeq 0.5\, a_X$.
So, for $d$ small, $W=-\xi^{exch}_d$ is positive and
%$-\xi_d (_{0 0}^{0 0})>0$,
the lowest energy state is the unpolarized state $|\bar\Psi_{N,N}\rangle$, while for $d>d^*$, the exchange Coulomb scattering turns positive and the lowest energy state is fully polarized, either 
${|\bar\Psi}_{2N,0}\rangle$ or ${|\bar\Psi}_{0,2N}\rangle$. Remarkably, we note that most experimental works are performed in a regime where $d\sim$ 10 nm \cite{Muljarov_2012}, i.e. in the range of the threshold distance $d^*$ for the switch of the condensate polarisation. This switch induced by fermion exchanges strikingly contrasts with recent theoretical studies \cite{Malpuech_2013,Kavokin_2012,Kavokin_2013,Hakioglu_2009} that have emphasized the role played by spin-orbit interactions in Bose-Einstein condensation of excitons.

\textbf{Conclusion }: We have shown that fermion exchanges between excitons may lead to a "spin-fragmented" condensate, \textit{i.e.}, a condensate in which all excitons do not have the same spin. For spatially indirect or "dipolar" excitons, we have then found that spin polarization has to be carefully considered in the context of Bose-Einstein condensation. Precisely, we predict that unpolarized and linearly polarized condensates have quasi-degenerate energies when the layers confining electrons and holes have a spatial separation $d$ below a threshold, $d^*$$\approx$$a_X$/2. On the other hand, for $d> d^*$ a drastic switch occurs towards a fully polarized condensate in which all excitons are in the same state, including spin - as in standard Bose-Einstein condensation. Our findings may then well relate to the low degree of polarisation observed in recent experiments reporting a condensate of dark excitons \cite{francois} and where $d\sim d^*$. Let us finally note that naturally our calculation is for a model situation where the quantum wells confining electronic carriers are considered purely 2D. Nevertheless we expect the corresponding result for actual experimental studies to fall in the same range.

F.D. acknowledges support from the European projects (EU) ITN-INDEX and EU-CIG XBEC.

\end{document}